\documentclass[sigconf]{acmart}
\usepackage[ruled,vlined,linesnumbered]{algorithm2e}

\AtBeginDocument{%
  }

\copyrightyear{2026}
\acmYear{2026}
\setcopyright{cc}
\setcctype{by}
\acmConference[ICSE '26]{2026 IEEE/ACM 48th International Conference on Software Engineering}{April 12--18, 2026}{Rio de Janeiro, Brazil}
\acmBooktitle{2026 IEEE/ACM 48th International Conference on Software Engineering (ICSE '26), April 12--18, 2026, Rio de Janeiro, Brazil}
\acmPrice{}
\acmDOI{10.1145/3744916.3773221}
\acmISBN{979-8-4007-2025-3/26/04}

\begin{document}

\title{MaCTG: Multi-Agent Collaborative Thought Graph for Automatic Programming}

\author{Zixiao Zhao}
\affiliation{%
  \institution{University of Auckland}
  \city{Auckland}
  \country{New Zealand}
} 
\email{zixiao.zhao@auckland.ac.nz}

\author{Jing Sun}
\affiliation{%
  \institution{University of Auckland}
  \city{Auckland}
  \country{New Zealand}
}
\email{jing.sun@auckland.ac.nz}

\author{Zhe Hou}
\affiliation{%
  \institution{Griffith University}
  \city{Brisbane}
  \country{Australia}
}
\email{z.hou@griffith.edu.au}

\author{Zhiyuan Wei}
\affiliation{%
 \institution{Beijing Institute of Technology}
 \city{Beijing}
 \country{China}
 }
 \email{weizhiyuan@bit.edu.cn}

\author{Cheng-Hao Cai}
\affiliation{%
  \institution{Suzhou Industrial Park Monash Research Institute of Science and Technology}
  \city{Suzhou}
  \state{Jiangsu}
  \country{China}
  }
\email{Cheng-Hao.Cai@monash.edu}

\author{Miao Qiao}
\affiliation{%
  \institution{University of Auckland}
  \city{Auckland}
  \country{New Zealand}
  }
\email{miao.qiao@auckland.ac.nz}

\author{Jin Song Dong}
\affiliation{%
  \institution{National University of Singapore}
  \country{Singapore}
}
\email{dcsdjs@nus.edu.sg}

\renewcommand{\shortauthors}{Zhao et al.}

\begin{abstract}
  With the rapid advancement of Large Language Models (LLMs), LLM-based approaches have demonstrated strong problem-solving capabilities across various domains. However, in automatic programming, a single LLM is typically limited to function-level code generation, while multi-agent systems composed of multiple LLMs often suffer from inefficient task planning. This lack of structured coordination can lead to cascading hallucinations, where accumulated errors across agents result in suboptimal workflows and excessive computational costs. 
  To overcome these challenges, we introduce MaCTG (Multi-Agent Collaborative Thought Graph), a novel multi-agent framework that employs a dynamic graph structure to facilitate precise task allocation and controlled collaboration among LLM agents. MaCTG autonomously assigns agent roles based on programming requirements, dynamically refines task distribution through context-aware adjustments, and systematically verifies and integrates project-level code, effectively reducing hallucination errors and improving overall accuracy. 
  MaCTG enhances cost-effectiveness by implementing a hybrid LLM deployment, where proprietary models handle complex reasoning, while open-source models are used for routine coding and validation tasks.
  To evaluate MaCTG’s effectiveness, we applied it to traditional image processing auto-programming tasks, achieving a state-of-the-art accuracy of 94.44\%. Additionally, by leveraging its hybrid LLM configuration, MaCTG significantly reduced operational costs by 89.09\% compared to existing multi-agent frameworks, demonstrating its efficiency, scalability, and real-world applicability. 
  Our project can be found at \url{https://github.com/MaCTG2025/MaCTG}.
\end{abstract}




\maketitle

\section{Introduction}
Since the release of OpenAI’s ChatGPT service in 2022~\cite{chatgpt}, research on Large Language Models (LLMs) has advanced rapidly, leading to the development of various frameworks that demonstrate fundamental problem-solving capabilities~\cite{yao2023treethoughtsdeliberateproblem, orru2023human, wu2024mathchat}. In the field of software engineering, numerous LLM-based studies and products have shown that, when trained on large-scale programming datasets, LLMs can learn both the syntax of programming languages and the logical structure of code, enabling them to perform preliminary code generation~\cite{chen2022codet, wang2023codet5plus, zhang2023planning}. However, LLMs remain highly constrained when generating large-scale, complex software systems. Current models are primarily limited to function-level code generation and struggle with project-level software development, where multiple interdependent components must be designed, implemented, and integrated seamlessly~\cite{du2024evaluating}. A major limitation of LLMs in large-scale software generation is their inability to effectively plan, coordinate, and verify code consistency across an entire project. LLMs operate on limited context windows, making it difficult to maintain coherence between different parts of a large software system. Furthermore, without structured high-level task planning and iterative refinement, LLMs often produce inconsistent, redundant, or incompatible code segments, requiring extensive human intervention to ensure correctness.

To address these challenges, some studies have proposed multi-agent systems where multiple LLMs collaborate by taking on specialised roles—such as project planning, code implementation, testing, and documentation—to tackle complex software projects~\cite{wang2024survey, qian2024chatdev, hong2023metagpt}. However, these top-down cascading approaches often oversimplify complex software engineering tasks. Individual agents are still burdened with single-level, isolated responsibilities, leading to cascading hallucinations, where errors accumulate throughout the hierarchy due to the lack of global oversight and verification mechanisms. Moreover, existing multi-agent frameworks frequently rely exclusively on proprietary LLMs, further exacerbating cost constraints and limiting accessibility. The lack of open, adaptable models and the prolonged iterative debugging process significantly increase the resource demands of these tools, making them impractical for scalable, real-world software development. Addressing these critical limitations requires a structured, adaptive, and cost-effective approach to LLM-driven software engineering.

In this paper, we introduce the Multi-agent Collaborative Thought Graph (MaCTG), an advanced multi-agent framework built on a dynamic graph structure to enhance auto-programming capabilities. Unlike traditional multi-agent systems with static roles and fixed task assignments, MaCTG dynamically determines the number and roles of agents required for a given programming task, ensuring flexibility and adaptability. To optimise workload distribution, MaCTG employs a two-dimensional task decomposition approach. First, horizontal decomposition breaks tasks into smaller, manageable components, reducing the burden on individual agents. Second, vertical cascading structures agents in a level-oriented hierarchy, ensuring logical task execution and systematic code generation. For effective agent communication, MaCTG integrates a context-aware refinement mechanism, inspired by graph-of-thought reasoning~\cite{besta2024graph}, to dynamically adjust task planning and coordination. This ensures precise control over each agent’s responsibilities, minimising redundancy and improving overall efficiency.
During the implementation phase, MaCTG applies a backwards validation workflow, which iteratively verifies and refines agent outputs, ultimately leading to the seamless assembly of project-level code. This structured approach—dynamic agent configuration, adaptive task planning, and multi-level verification—effectively mitigates cascading hallucinations, enhancing software reliability and accuracy. Additionally, MaCTG employs a hybrid model configuration, integrating proprietary models (DeepSeek-V3~\cite{deepseekai2024deepseekv3technicalreport}) with a locally deployed open-source model (Qwen2.5-Coder-7B~\cite{qwen2.5}). This combination optimises performance and cost-efficiency, ensuring scalability while reducing reliance on proprietary models. By balancing high-level reasoning and routine coding tasks, MaCTG offers a practical and cost-effective solution for large-scale software development.

To assess the effectiveness of MaCTG in automatic programming for complex projects, we selected a traditional image-processing scenario as a case study. 
For this evaluation, we developed the BCVPP dataset, comprising 90 projects across three levels of automatic programming tasks. This dataset provides a comprehensive benchmark for testing the capabilities of auto-programming tools in real-world applications. Using the BCVPP dataset, we conducted a detailed comparative analysis of open-source LLMs, proprietary LLMs (and tools), and other multi-agent systems. MaCTG achieved an execution accuracy of 94.44\%, surpassing competing approaches. Furthermore, by leveraging its hybrid LLM configuration, MaCTG reduced operational costs by 89.09\% compared to existing multi-agent frameworks, highlighting its efficiency, scalability, and practical applicability in software development. 

The key contributions of this work are as follows: 
1) We propose MaCTG, a novel multi-agent framework utilising a dynamic graph structure with context-aware planning adjustment and multi-scale code refinement, enabling more effective collaboration for complex automatic programming tasks.
2) We introduce a hybrid LLM configuration combining proprietary models for complex reasoning with open-source models for routine tasks, significantly improving cost-efficiency while maintaining performance.
3) We develop BCVPP, a benchmark dataset for evaluating automatic programming frameworks in image processing, and conduct a detailed comparative analysis of LLM-based frameworks in complex project-level tasks.

The remainder of this paper is organised as follows. 
Section 2 reviews related work on automatic programming and existing approaches.
Section 3 details the workflow of MaCTG, including its multi-agent collaborative thought graph structure and multi-scale validation and assembly process. 
Section 4 describes the experimental setup, covering the BCVPP dataset construction and the evaluation of MaCTG’s performance against state-of-the-art methods. 
Section 5 concludes by summarising key contributions and proposing future research directions.
\begin{figure*}[htbp]
  \centering
  \includegraphics[width=\linewidth]{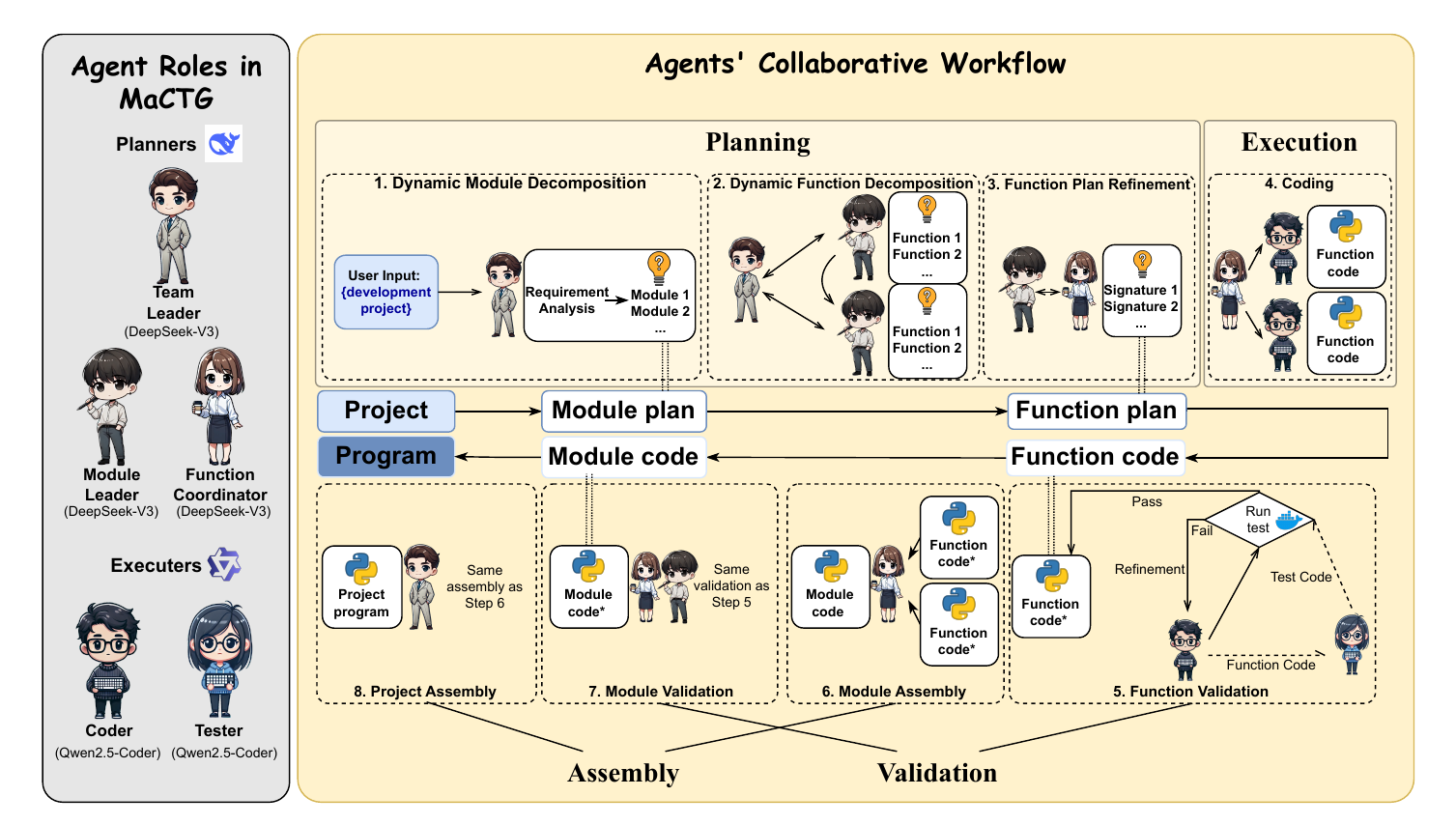}
  \caption{MaCTG’s collaborative workflow of four phases: Planning, Execution, Validation and Assembly}
  \Description{brief workflow of MaCTG}
  \label{fig:workflow}
\end{figure*}

\section{Related work}
In this section, we provide background on the auto-programming tasks and methods, particularly agent-based solutions.

\subsection{Automatic programming}
The concept of automatic programming dates back to the 1950s when it was primarily considered a compilation problem — translating formal specifications into implementations~\cite{balzer198515}. 
Over time, this evolved into generating domain-specific languages from task descriptions~\cite{kushman2013using,raza2015compositional,manshadi2013integrating}. 
By the 2010s, advancements in natural language processing (NLP) based on deep learning introduced algorithms leveraging RNNs and LSTMs~\cite{yin2017syntactic}, enabling initial natural language to programming language translation.
The Transformer architecture~\cite{vaswani2017attention} revolutionized this field, spurring development of tools like CodeBERT~\cite{feng2020codebert} and establishing foundations for LLMs such as CodeLlama~\cite{roziere2023code} and StarCoder~\cite{li2023starcoder}. 
The increasing accuracy and scalability of these models have enabled applications extending beyond code generation, including program repair and bug detection, as demonstrated by  LLM4PR~\cite{cai2024towards} and InterFix~\cite{jin2023inferfix}.

Despite these advancements, the ability of a single LLM to tackle complex project-level programming tasks remains constrained. 
These tasks require nuanced understanding, effective task planning and integration across modules—capabilities that current LLMs struggle to achieve due to limitations in the scope of training data and architectural design. 
This gap has motivated the exploration of agent-based frameworks and advanced prompt engineering techniques to extend the potential of LLMs in automatic programming.

\subsection{Agent-based solutions in auto-programming}
The code that LLMs can generate through a single round of dialogue is often relatively simple and lacks the complexity required for auto-programming tasks. 
To address this limitation, frameworks have been developed that use an LLM as an AI agent, enhanced with specific interaction strategies~\cite{wang2024survey, xi2023rise}. 
For instance, ToolCoder~\cite{zhang2023toolcoder} and CodeAgent~\cite{zhang2024codeagent} enhance LLM capabilities by integrating API search, format checkers, and code interpreters. 

Beyond individual agent enhancement, researchers have proposed multi-agent systems for collaborative programming~\cite{wang2024survey}.
CAMEL~\cite{li2023camel} provides basic Q\&A communication mechanisms among agent roles. 
ChatDev~\cite{qian2024chatdev} assigns distinct roles to agents with cascading conversation strategies. AgentCoder~\cite{huang2023agentcoder} enhances testing agent functionality, while MetaGPT~\cite{hong2023metagpt} integrates SOPs from human software engineering. 
However, these frameworks rely on rigid, vertical stage-level task division with limited collaboration, resulting in restricted modularity and adaptability.
The recent Manus framework~\cite{manus} assigns agents to detailed stages for end-to-end problem-solving; however, it was not publicly available for evaluation during this study.
Other approaches like SWE-Agent~\cite{yang2024swe} and AutoCodeRover~\cite{zhang2024autocoderover} focus on bug fixing rather than modular code generation, while OpenAgents~\cite{xie2023openagents} targets LLM tool invocation rather than collaborative programming.
Despite these advancements, most existing multi-agent systems operate sequentially with one active agent per stage, placing high demands on individual agents and limiting effectiveness in complex task management.

\section{Methodology}
In this section, we detail the structure and workflow of MaCTG.  
Section~\ref{sec:agent} describes the multi-agent configuration, Section~\ref{sec:planning} focuses on the structured planning workflow and context-aware adjustments, and Section~\ref{sec:backward} covers the multi-level validation and assembly, emphasising their roles in ensuring output reliability.

The overall architecture of MaCTG is illustrated in Figure~\ref{fig:workflow}, comprising four distinct phases: \textbf{Planning, Execution, Validation, and Assembly}. 
The \emph{Planning} phase begins with the Team Leader, who analyses project requirements, decomposes tasks into modules, and assigns them to Module Leaders. Instead of working in isolation, Module Leaders collaborate closely, refining task scopes and ensuring cross-module consistency under the Team Leader’s supervision. Through structured context-sharing, they align dependencies and interactions before passing refined function plans to Function Coordinators. 
Function Coordinators further structure these plans, generating precise function signatures while maintaining synchronisation with Module Leaders. In the \emph{Execution} phase, Coders implement the code according to these specifications, continuously interacting with Function Coordinators to ensure correctness and adaptability. 
The \emph{Validation} phase involves Testers, who work closely with Coders to develop and execute test cases. Errors are identified and iteratively refined through a structured feedback loop, where Coders, guided by Function Coordinators, resolve issues before advancing to the \emph{Assembly} phase. 
During assembly, validated functions are integrated into complete modules by Function Coordinators, followed by additional validation before final integration by the Team Leader. Throughout this structured workflow, real-time collaboration between agents ensures adaptability, minimises errors, and maintains overall system integrity. 
MaCTG’s multi-agent collaboration mechanism dynamically adjusts code generation strategies, optimises expertise distribution, and enhances project-level automation through seamless agent interaction. 
By integrating structured planning, iterative validation, and coordinated execution, MaCTG delivers scalable, reliable, and high-quality software development.

\subsection{Agents in MaCTG framework}
\label{sec:agent}
\subsubsection{Agent specification}
MaCTG incorporates five distinct types of agents: Team Leader, Module Leader, Function Coordinator, Coder, and Tester.
The first three roles—Team Leader, Module Leader, and Function Coordinator—are collectively referred to as \textbf{Planners}, as they are responsible for task decomposition, high-level validation, and structured assembly within the workflow. 
In contrast, the Coder and Tester, known as \textbf{Executors}, focus solely on function-level implementation and validation. 

Providing clear and unambiguous definitions for each agent is essential to ensure their understanding of their roles within the overall workflow, particularly when tackling complex tasks such as automatic programming~\cite{hong2023metagpt}. 
To define the Planner roles, we followed a structured approach by defining each role’s hierarchical placement first, then specifying their interactions with superior and subordinate agents, simultaneously outlining their input, output, and execution scope.
Additionally, each role also has unique specifications to avoid errors and enhance performance.
For \emph{Team Leader}, we prompt it to minimise the number of modules during Planning to minimise complexity and redundancy. 
In the Assembly phase, the Team Leader is prompted to reference existing modules rather than redefining them, reducing duplication and token usage.
\emph{Module Leader} manages function decomposition within a module.
We require it to ensure logical coherence and execution order to facilitate seamless plan propagation to the Function Coordinator.
\emph{Function Coordinator} generates function signatures based on function plans. 
It is demanded to include explicit dependency declarations to streamline implementation and prevent missing imports.
When correcting function code during the validation phase, instructions for Coders are required to be structured item-by-item rather than in unstructured descriptions, improving correction accuracy.

For the Executor roles, definitions focus primarily on coding constraints:
\emph{Coders} must strictly adhere to declared dependencies in function signatures to prevent missing imports or broken module references.
\emph{Testers} must use the unittest module for function validation, correctly importing tested functions while ensuring cleanup operations do not delete essential source files (e.g., input images). 
By establishing clear specifications for both Planners and Executors, the framework ensures seamless collaboration among agents, enhancing the efficiency and effectiveness of the overall workflow.

\subsubsection{Hybrid agent employment}
Another critical aspect of agent configuration is the selection of the base model. 
Proprietary models, such as Claude~\cite{claude3.7},  GPT~\cite{achiam2023gpt} and DeepSeek~\cite{deepseekai2025deepseekr1incentivizingreasoningcapability}\footnote{DeepSeek R1 and V3, while open-source, are categorized as proprietary models in this study due to their large-scale architecture, which makes local deployment impractical, necessitating API-based access.}, typically exhibit superior comprehension and generation capabilities due to their larger scale and broader training datasets. 
These attributes make proprietary models particularly suitable for managing overarching tasks that require high-level reasoning and adaptability~\cite{jiang2024survey}. 
In contrast, open-source LLMs, while generally smaller in scale and trained on more domain-specific datasets, are well-suited for handling smaller, context-specific tasks with targeted precision.

To capitalise on these complementary strengths, MaCTG adopts a hybrid model strategy. 
In our implementation, we utilise DeepSeek-V3 as the base model for the Planner roles, and the open-source model Qwen2.5-Coder-7B~\cite{qwen2.5} for the Executors. 
This configuration is designed to balance performance and cost-effectiveness with the following rationale:
\textbf{Planners: }DeepSeek-V3 was selected due to its state-of-the-art capabilities and is the most recently released model at the time of this study. Its advanced reasoning ability ensures effective task decomposition, structured planning, and inter-agent coordination. Meanwhile, the competitive pricing compared with other proprietary models makes it ideal for MaCTG's implementation.
\textbf{Executors: }The Coder and Tester roles focus exclusively on detailed programming tasks, guided by highly specific and well-structured prompts. As such, the capability requirements for these roles are comparatively lower. For these tasks, we utilise Qwen2.5-Coder, a novel open-source model optimised for code generation, providing adequate performance without any API cost.
	
By combining proprietary models for high-level tasks with open-source models for more granular operations, this approach ensures that the MaCTG framework maintains high performance while simultaneously reducing overall system costs. 

\subsubsection{Dynamic multi-agent structure}
\begin{figure}
    \centering
    \includegraphics[width=0.9\linewidth]{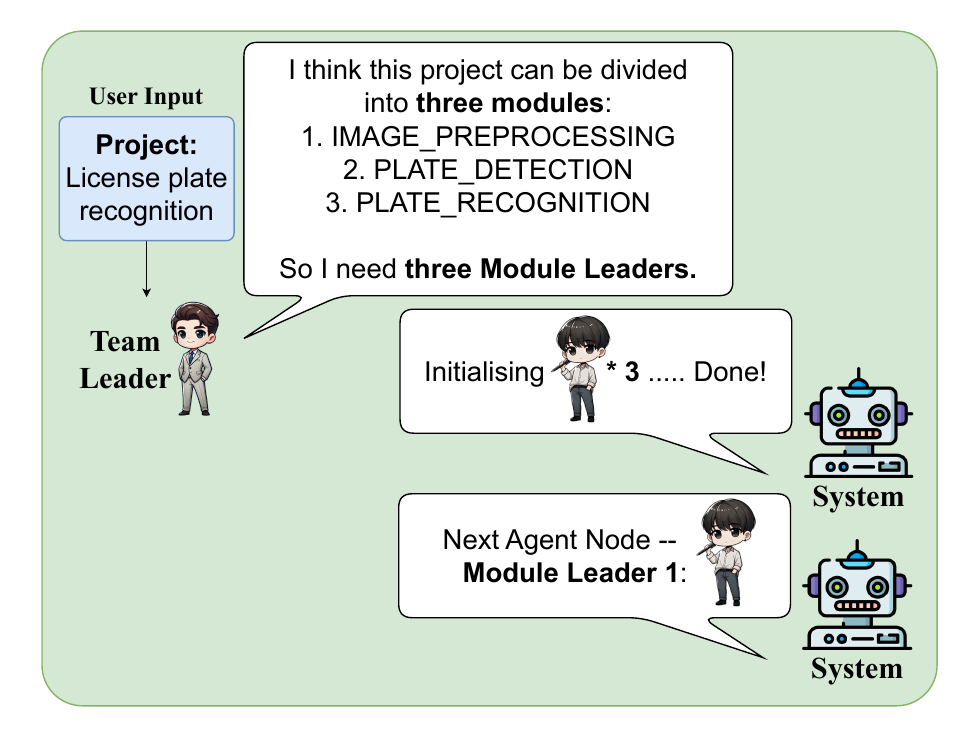}
    \caption{Dynamic Module Leader configuration: Module Leaders are only initialised after the Team Leader determines the module decomposition.}
    \label{fig:dynamic_decompo}
    \Description{decomposition}
\end{figure}

\begin{figure*}[h!]
    \centering
    \includegraphics[width=0.9\linewidth]{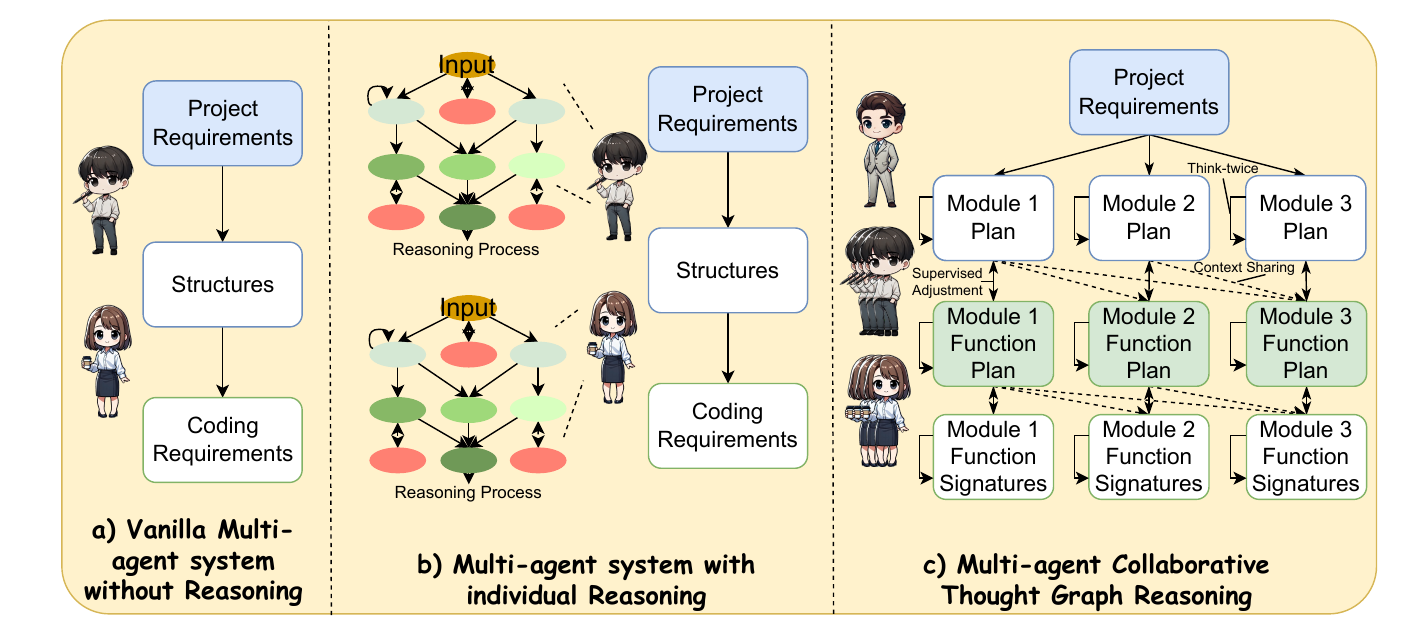}
    \caption{Comparison of Multi-agent planning process}
    \label{fig:reasoning}
    \Description{Comparison of Path Reasoning process}
\end{figure*}
In the Plan \& Execute workflow, agents are dynamically initialised as needed. 
Figure~\ref{fig:dynamic_decompo} illustrates an example configuration of the Module Leaders. 
At the start of the project, only the Team Leader exists as an agent in MaCTG. 
The Team Leader performs the initial module planning based on the user’s input. 
Once the module plan is finalised, the Module Leaders are initialised and added to the agent graph. 
This dynamic initialisation process continues at subsequent levels; Coders are activated to implement functions only after the Function Coordinator has defined the function signatures.

The adoption of horizontal, multi-agent task decomposition at each planning step—rather than relying on a single agent, as seen in other multi-agent frameworks—addresses the issue of self-consistency in an individual agent’s thought process. 
In a single-agent process, hallucinations in one part of a plan can cascade, adversely affecting the entire workflow. 
In contrast, MaCTG’s design allows peer agents to independently re-examine and validate plans at every level, introducing opportunities for verification and correction. 
This collaborative structure effectively mitigates the risk of cascading hallucinations, enhancing the overall robustness and reliability of the framework.

\subsection{Collaborative graph reasoning protocol}
\label{sec:planning}
\subsubsection{Multi-agent collaborative thought graph}
Recent advances in structured reasoning include Chain of Thought (CoT)~\cite{wei2022chain}, Tree of Thoughts (ToT)~\cite{yao2023treethoughtsdeliberateproblem}, and Graph of Thoughts (GoT)~\cite{besta2024graph}, which introduced dynamic optimisation across reasoning paths. 
Inspired by these strategies, we aim to integrate structured reasoning into multi-agent workflows to enhance task coordination and execution.

Classical multi-agent systems adopt a cascaded structure (Figure~\ref{fig:reasoning}a), where LLM-driven agents lack reasoning capabilities and rely solely on prompts to advance tasks.
An improvement is to incorporate reasoning into each agent (Figure~\ref{fig:reasoning}b), allowing individual agents to autonomously generate path reasoning and derive their respective results.
However, this local reasoning approach has a critical limitation: agents prone to overthinking may unnecessarily split straightforward programming tasks into overly subtle sub-functions, increasing dependencies and ultimately amplifying implementation difficulties and hallucinations.

To mitigate these issues, MaCTG restructures reasoning as a collaborative process within an integrated graph-based multi-agent system.
As illustrated in Figure~\ref{fig:reasoning}c, MaCTG employs a directed graph structure where each node represents a discrete thought process handled by a separate agent.
Rather than assigning reasoning tasks to individual agents, reasoning MaCTG emerges from structured inter-agent collaboration through directed edges that facilitate: 
(1) an agent’s think-twice process for internal extension,
(2) context sharing among peer agents to maintain coherence, and
(3) supervised adjustment through leader-subordinate interactions to ensure overall consistency.
By distributing reasoning across all agents and limiting each agent’s focus to a well-defined subtask, MaCTG significantly reduces cognitive overload and prevents overthinking.
This collaborative approach simplifies each agent’s decision-making, minimising redundant decomposition while effectively reducing hallucinations and improving execution reliability.

To formalise our approach, we use the following definitions to better illustrate CTG:

\textbf{Definition 1 (Collaborative Thought Graph):} A CTG is a directed acyclic graph $G = (V, E, C, \Phi)$ where:
\begin{itemize}
\item $V = \{v_i^j | i \in \text{levels}, j \in \text{agents at level } i\}$ represents the set of agent nodes, where each $v_i^j$ corresponds to a specific agent at level $i$ 
;
\item $E \subseteq V \times V$ represents directed dependency relationships between agents, capturing both hierarchical supervision and peer collaboration;
\item $C: V \rightarrow \mathcal{S}$ maps each agent node to its current context state, where $\mathcal{S}$ denotes the context space containing task specifications, intermediate results, and shared knowledge;
\item $\Phi: V \times \mathcal{S} \rightarrow \mathcal{S}$ represents context update functions that define how agents process contextual information.
\end{itemize}

\textbf{Definition 2 (Context Propagation):} For an agent $v_i^j \in V$, the context update follows:
\begin{equation}
C(v_i^j) = \Phi(v_i^j, \mathcal{A}(\{C(v_k^l) | (v_k^l, v_i^j) \in E\}))
\end{equation}
where $\mathcal{A}$ is an aggregation function that combines contexts from predecessor agents, and $\Phi(v_i^j, \cdot)$ represents the agent-specific context processing that incorporates both received context and the agent's local thoughts.
In practice, the aggregation function $\mathcal{A}$ is implemented as a structured dictionary that organises peer agents' plans by module names and descriptions, enabling systematic access to relevant contextual information. 
The context update function $\Phi$ is operationalised through agent-specific prompt templates that instruct the LLM on how to integrate the aggregated context with the agent's current task.

Building upon this formal framework, we now detail how MaCTG implements the three key coordination mechanisms through structured planning workflows and context-aware adjustments.

\subsubsection{Two-step planning and potential planning error}
\begin{figure}
    \centering
    \includegraphics[width=0.9\linewidth]{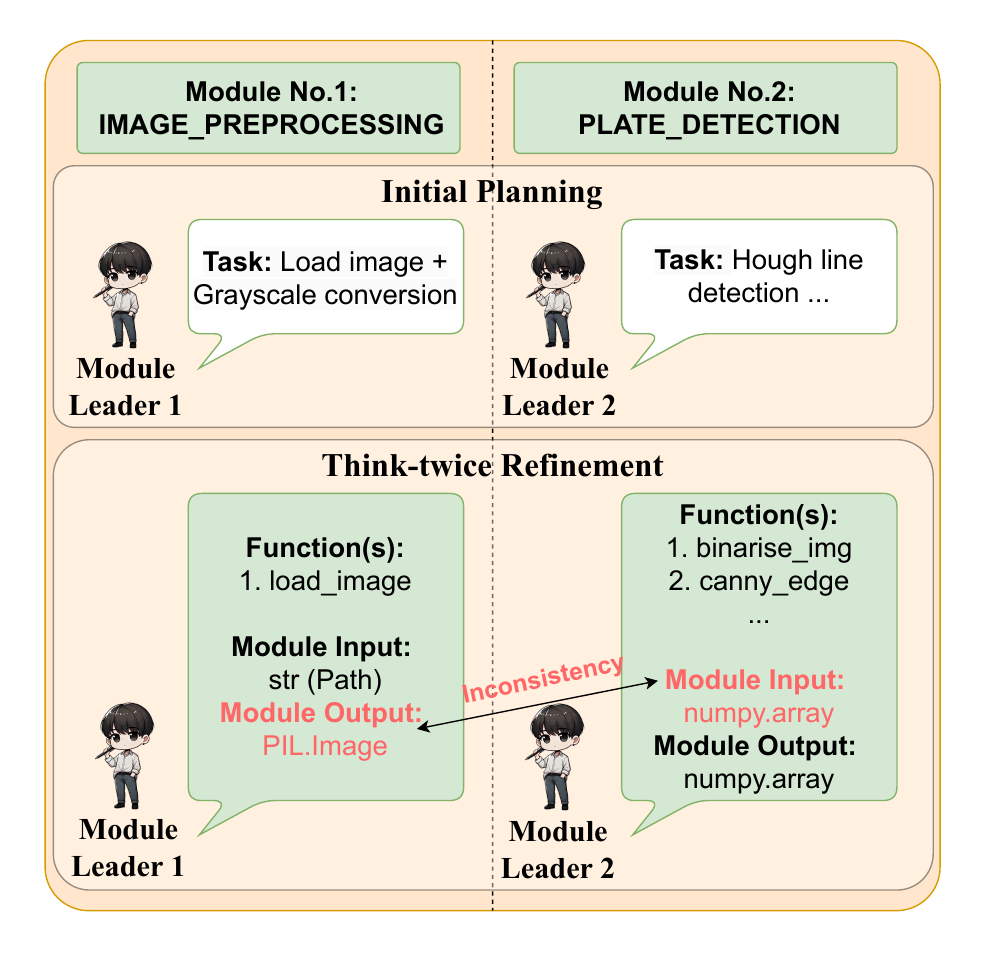}
    \vspace{-8pt}
    \caption{Agent planning error: Separate planning procedure leads to inconsistencies between modules.}
    \label{fig:init_plan}
    \vspace{-8pt}
    \Description{planning error}
\end{figure}

In the actual task flow, following the principles of ReAct~\cite{yao2022react}, we decompose the planning process of an agent into two sequential steps: initial planning and think-twice extension.

During the initial planning phase, each agent at level $i$ is assigned a subset of the overall task set $T_i$, denoted as $T_i^j$ for agent $v_i^j$. The agent then generates a preliminary plan based on its tasks:
\begin{equation}
    P_i^{j(0)} = F_{\text{init}}(T_i^j)
\end{equation}
where \( F_{\text{init}} \) represents the agent's own reasoning process used to generate a preliminary plan.
Then, in the think-twice extension phase, each agent extends its preliminary plan into a structured and comprehensive form before passing it to the next level:
\begin{align}
C^{(1)}(v_i^j) &= \Phi_{think}(v_i^j, C^{(0)}(v_i^j)) \\
P_i^{j(1')} &= F_{\text{ext}}(P_i^{j(0)}, C^{(1)}(v_i^j))
\end{align}
where $C^{(0)}(v_i^j)$ contains the initial task assignment $T_i^j$, $\Phi_{think}$ updates the context through internal reasoning, and $F_{\text{ext}}$ generates the final plan using the enriched context, ensuring the generated plan follows the expected format and maintains logical coherence.

However, despite this two-step reasoning process, a fundamental limitation remains if agents operate in isolation: an agent's extended plan depends only on its assigned task subset and local context, without considering dependencies across agents at the same level: 
\begin{equation}
    P_i^{j(1')} = F_{\text{ext}}(F_{\text{init}}(T_i^j), C_{local}(v_i^j)), \quad \text{where } T_i^j \perp T_i^k, \forall k \neq j
\end{equation}
where $C_{local}(v_i^j)$ represents isolated context without peer information, potentially leading to contextual inconsistencies across different plans.
These inconsistencies manifest in two key ways: discrepancies in intermediate variable types across plans and overlapping or redundant function plans among multiple agents working in parallel.
As illustrated in Figure~\ref{fig:init_plan}, taking the Team Leader–Module Leader structure as an example, after the Team Leader defines the scope of each module, the module descriptions are distributed individually to multiple Module Leaders for further planning. 
Since each module leader only focuses on its own assigned module, this independent execution may create hallucinations in the generated plan, leading to mismatches in the definition of intermediate variables and ultimately making the generated program unusable. 

To address these issues, we introduce a context-aware planning adjustment strategy, which enhances inter-agent coordination and ensures consistency across plans.

\subsubsection{Context-aware planning adjustment}
The purpose of context-aware planning adjustment is to provide each agent with a comprehensive view of the evolving context, thereby enabling dynamic revision of its preliminary plan and resolution of potential inconsistencies. 
This methodology mirrors real-world software development workflows, in which each subgroup devises an initial plan and iteratively refines it based on the contributions of peer subgroups.

In MaCTG, this process is realised through two sequential steps: \emph{context sharing} and \emph{supervised adjustment}.

\textbf{Context sharing:} The context-sharing mechanism facilitates communication among agents at the same hierarchical level.
After an agent generates its initial plan, it receives the confirmed plans of its peers during the "think-twice" process, thereby obtaining a comprehensive understanding of the overall context. 
Mathematically, this is expressed as:
\begin{align}
C_{peers}(v_i^j) &= \mathcal{A}_{peer}(\{C^{(1)}(v_i^k) | k \neq j, v_i^k \in \text{same level}\}) \\
C^{(2)}(v_i^j) &= \Phi_{context}(v_i^j, C^{(1)}(v_i^j) \cup C_{peers}(v_i^j))
\end{align}
where $\mathcal{A}_{peer}$ aggregates contextual information from peer agents, and $\Phi_{context}$ updates the agent's context by incorporating peer insights to resolve interface inconsistencies and avoid redundant functionality.

Then, the extended plan of agent \( v_i^j \) at level \( i \) is corrected as:
\begin{equation}
    P_i^{j(1)} = F_{\text{ext}}\left(P_i^{j(0)}, C^{(2)}(v_i^j)\right)
\end{equation}
\\[1pt]
\textbf{Supervised adjustment:} Supervised adjustment requires interaction between an agent and its superior agent.
Upon finalising its plan, an agent submits its output to its leader for validation and further adjustment:
\begin{align}
C_{leader}(v_i^j) &= \mathcal{A}_{hierarchical}(C^{(2)}(v_{i-1}^{parent(j)})) \\
C^{(final)}(v_i^j) &= \Phi_{supervise}(v_i^j, C^{(2)}(v_i^j) \cup C_{leader}(v_i^j))
\end{align}
\\[1pt]
 where $v_{i-1}^{parent(j)}$ represents the leader agent, $\mathcal{A}_{hierarchical}$ extracts relevant guidance from the leader's context, and $\Phi_{supervise}$ ensures alignment with global objectives and corrects lower-level inconsistencies.

Finally, the supervisor-adjusted plan is expressed as:
\begin{equation}
    P_i^{j(final)} = F_{\text{sup\_adjust}}\left(P_i^{j(1)}, C^{(final)}(v_i^j)\right)
\end{equation}




\begin{figure}
    \centering
    \includegraphics[width=0.9\linewidth]{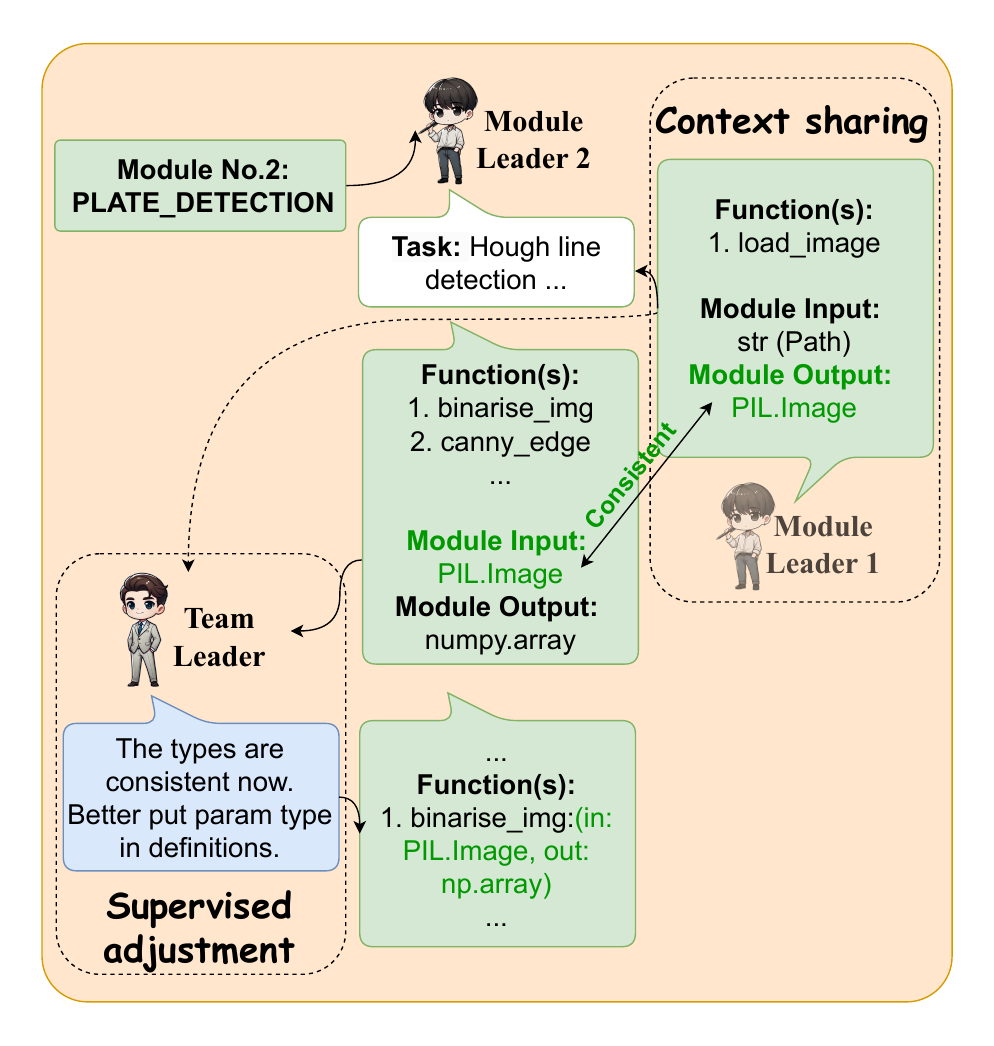}
    \caption{Context-aware planning adjustment: Context sharing mitigates the inter-module inconsistency; Supervised adjustment further reinforces structured communication.}
    \label{fig:context_adjust}
\end{figure}

Figure~\ref{fig:context_adjust} illustrates how context-aware planning adjustment addresses inconsistencies in intermediate variable types between modules.
By obtaining a clear understanding of the output types from preceding modules through context sharing, the current Module Leader can adhere to these contextual requirements during its "think twice" process, thereby avoiding variable-type conflicts. 
Additionally, supervised adjustment enables the Team Leader to enforce stringent standards on function specifications, providing a more coherent plan for subsequent developmental stages.

In automatic programming tasks, the quality of the prompt is a critical factor influencing the accuracy and reliability of the generated code~\cite{khojah2024impact}. 
Through the two-step context-aware planning adjustment process applied to each planner agent, MaCTG systematically translates natural language project requirements into coherent function signatures, which serve as structured prompts for coder agents during the implementation phase. 
By predefining function inputs, outputs, and dependencies with precision, this approach enhances code correctness and minimises inconsistencies in execution. 
This planning strategy ensures that individual agents operate within a well-structured framework, where context sharing mitigates hallucinations, and supervised adjustment further clarifies the plan before execution, leading to a more reliable and efficient automatic programming workflow.

\subsection{Multi-scale code validation and assembly}
\label{sec:backward}
For automatic programming tasks, the essence of program quality lies in the accuracy of the code.
In addition to enhancing code requirement descriptions during the Plan \& Execute process, we have designed a multi-scale validation and assembly strategy to fix code vulnerabilities and prevent the propagation of errors.

\textbf{Function code validation and module assembly: } After a well-structured planning process, Coder agents implement the lowest-level function code.
At the finest granularity of the system, these functions must be rigorously validated before being integrated into larger modules.
To ensure correctness, a Tester agent—guided by the Function coordinator—generates and executes test cases tailored to the function code.
During validation, both the function and test code are mounted into a predefined Docker container preconfigured with all dependencies required for image processing in Python. 
Test execution follows a fixed command, ensuring standardised output collection. If a failure occurs, the traceback output pinpoints the exact line and cause of the error.
The Function Coordinator then analyses the test results, diagnosing potential issues in the implementation.  
If necessary, it provides refinement guidance to the Coder, ensuring failures are addressed iteratively.
Once all assigned functions are validated, the Function Coordinator aggregates them into a cohesive module, ensuring structural consistency before propagating upward for higher-level validation.

\textbf{Module validation and project assembly: }
After the Module Leader receives the assembled module code, it then validates the module through a higher-scope test code, checking for inconsistencies in function dependencies, and modular data flow.
The overall module validation process is consistent with that of function validation.
If discrepancies arise, it then generates the modification ideas for the Function Coordinator to revise the module code before resubmitting it for approval.  
Following module validation, the Team Leader integrates all verified modules into the final project codebase, ensuring a seamless connection between modules.  

This multi-scale validation and assembly process progressively refines correctness, mitigates inconsistencies, and prevents structural flaws, leading to a functional and coherent program output.

\section{Evaluation}

To thoroughly investigate the effectiveness of MaCTG, we formulate the following research questions:

\textbf{RQ1: How does the dynamic graph-based multi-agent framework perform in auto-programming tasks?}  

\textbf{RQ2: How does the performance of MaCTG compare to other auto-programming methods?}  

\textbf{RQ3: How do context-aware planning adjustment and multi-level code validation reduce hallucinations and improve accuracy?}  

\textbf{RQ4: How does MaCTG’s collaborative reasoning strategy enhance performance compared to multi-agent systems with independent agent reasoning?}

\textbf{RQ5: How does a hybrid configuration of proprietary and open-source LLMs improve cost-efficiency?}  

\subsection{Benchmarking dataset}
We focus on project-level code generation rather than existing benchmarks like HumanEval~\cite{chen2021evaluating} and MBPP~\cite{austin2021program}, which target single-function generation, or SWE-bench~\cite{jimenez2024swebench}, which focuses on bug-fixing in existing codebases. 
These benchmarks do not evaluate the multi-agent collaboration and project-level planning capabilities that MaCTG is designed to address.
To our knowledge, no existing open datasets systematically evaluate auto-programming methods on complex, multi-module code generation tasks. 
To mitigate this gap, we compiled a test set of image-processing-related projects named BCVPP (Basic Computer Vision Python Programming). 
%
The primary objective of our project is to apply MaCTG to auto-programming in medical image processing, where precise and efficient code generation is essential for clinical applications~\footnote{While our initial focus is on medical imaging, the MaCTG framework is designed to be adaptable, allowing it to be extended to general-purpose code generation across various domains.}. At this stage, we have selected the natural image processing domain as a case study due to the availability of publicly accessible coding questions, enabling reproducible and comparative assessments.

BCVPP focuses on fundamental Python image processing projects, specifically within the scope of classical image algorithms. 
The dataset comprises a total of 90 projects, which are divided into 30 simple projects, 50 medium projects, and 10 hard projects, 
defined as follows: Simple projects provide explicit instructions within the project descriptions, clearly specifying what should be generated. 
Medium projects typically feature unclear connections between components, challenging methods to interpret inter-module dependencies and resolve integration issues autonomously. 
Hard projects present only high-level, vague requirements with minimal guidance, requiring methods to interpret necessary implementation details.
Below are some sample projects of categories inside BCVPP:\\
\textbf{Simple:} Given an RGB image ("./test\_image.png"), split it into three separate images, one for each channel (R, G, B) and save them as separate PNG files: "red\_channel.png", "green\_channel.png" and "blue\_channel.png". \textit{(This task provides explicit step-by-step instructions with clear channel split specifications.)}\\
\textbf{Medium:} Given an image ("./test\_image.png"), perform Canny edge detection (th=100, 200), identify the enclosed regions, and fill these regions with green while keeping the edges, then save the final image as "filled\_regions.png". \textit{(While individual operations are specified, the connections between edge detection, region identification, and color filling require interpretation of how to integrate these modules.)}\\
\textbf{Hard:} Given an input image ("./Phantom.png"), apply projection analysis to the image and save the output as "transformed.png". Then apply reconstruction to the transformed image and save the output as "reconstructed.png". \textit{(This task requires methods to interpret vague CT image processing and determine appropriate projection and reconstruction approaches without explicit guidance.)}

As a classic library for traditional image algorithms, numerous online courses and project repositories offer examples based on OpenCV.
After evaluating various sources, we selected the beginner-oriented GitHub repository "OpenCV\_Projects"~\cite{rchavezjOpenCV_Projects} as the primary source of test data. 
This repository, with over 340 stars and around 55 OpenCV-related problems, was chosen because its content is widely accepted by developers and offers an appropriate difficulty level.
We selected 30 projects from this repository, which are classified as \textbf{simple} class.

While simple projects are useful for initial validation, testing MaCTG’s performance on project-level tasks requires projects that involve the integration of multiple modules.
Unfortunately, such projects are typically found in real-world industrial scenarios and are rarely available in open-source repositories. 
To fill this gap, we utilised the OpenAI-o1~\cite{OpenAIo1} model to generate 100 moderately complex projects by combining elements from the 30 simple projects. 
However, during the calibration process, we discovered that many of these generated projects lacked coherent and logical relationships between the modules, leading to confusing project descriptions — for example, a case where a key-point detection module is followed by an unrelated histogram equalisation. 
After reviewing and filtering these projects, we finalised 50 \textbf{medium} projects. 

To further challenge the capabilities of MaCTG and other automatic programming methods, we sought to include more difficult projects. 
We enabled the “deep-research” tool in OpenAI-o1 to search for and generate 10 \textbf{hard} task ideas that push the limits in two key ways.
First, these projects are inherently more complex, requiring a greater number of modules and demanding a higher level of expertise. 
Second, unlike the clear instructions found in simpler projects, these projects offer only vague descriptions—typically providing just the input image and target requirements—forcing the methods to infer the necessary steps on their own.

In BCVPP, all requirements follow a standardised format: input image, expected operations, and expected outputs. 
For single-LLM baselines, we append the instruction “Write a set of Python code to finish the project.” 
For agent-based methods, including ours, only the raw requirement is provided, allowing each method to interpret and decompose the task autonomously. 
This setup ensures a fair and consistent comparison across all approaches.
\subsection{Evaluation metric \& configuration}
Our objective is to use the BCVPP dataset to evaluate various methods' ability to automate programming in the image processing domain. For each project $p$, given an input image $Img_p$ and processing requirement, each method generates algorithm code $\Phi_{project}$ to meet the requirement. 
The code is executed to produce output, and we verify whether it matches the expected outcome $Output_{exp}$. This process is represented by:
\begin{align}
    Output \gets \Phi_{project}(Img_p) \\
    Output_{exp} \gets \Phi_{sample}(Img_p) \\
    Acc_p = \begin{cases}
    1, & Output = Output_{exp}\\
    0, & Output \neq Output_{exp}	
           \end{cases}
\label{equ:passk}
\end{align}
The overall accuracy of each method on a $p$ is represented as $Acc_{p, i}$ for the $i$-th generated solution, with $Acc_{p, i} = 1$ if the output matches the expected outcome, and $0$ otherwise.
To determine $Acc_{p, i}$, we execute the generated code and compare its output to the ground-truth result. 
If the automated comparison fails but the generated code (1) executes successfully, (2) follows the task requirements, and (3) produces outputs of the expected type and format, we manually inspect the result to verify correctness, as differences may arise from valid library choices (e.g., OpenCV vs. Scikit-image) rather than implementation errors.
Summing across $n$ generated solutions gives $c_p = \sum_{i=1}^{n} Acc_{p, i}$, which is used in the \textbf{Pass@k} metric:
\begin{equation}
Pass@k = \mathop{\mathbb{E}}_{p \in \text{projects}} \left[ 1 - \frac{\binom{n - c_p}{k}}{\binom{n}{k}} \right]
\end{equation}
where $n$ is the total number of generated outputs per project and $k$ is the number of samples considered in Pass@k. To assess both initial problem-solving ability and reduce LLM output variance, we report \textbf{Pass@1} and \textbf{Pass@5} in our experiments.

The evaluation was performed on an Ubuntu 22.04 server equipped with an AMD Ryzen Threadripper PRO 3995WX CPU (64 cores), 252GB RAM, and 2 NVIDIA RTX A6000 GPUs (48GB VRAM each).

\subsection{Experimental results}
\begin{table*}[!htbp]
  \begin{center}
  \renewcommand\arraystretch{1.05}
  \setlength{\tabcolsep}{2.5pt}
  \caption{Comparison of auto-programming methodologies}
  \label{tab:comparison}
  \begin{tabular}{lcccc|ccccc}
    \hline
    & \multicolumn{4}{c|}{Pass@1} & \multicolumn{5}{c}{Pass@5} \\
    \cline{2-5} \cline{6-10}
      & Simple & Medium & Hard & Overall & Simple & Medium & Hard & Overall & Sig.\\
    \hline
    \textbf{Open-source models} & & & & & & & & & \\
    \hline
    \texttt{CodeLLaMA-Instruct-70B~\cite{roziere2023code}} & 30.00\% & 28.00\% & 20.00\% & 26.67\% & 40.00\% & 40.00\% & 20.00\% & 37.78\% & ***\\
    \texttt{LLaMA3.3-70B~\cite{llama3.3}}  & 70.00\% & 60.00\% & 30.00\% & 60.00\% & 76.67\% & 66.00\% & 40.00\% & 66.67\%  & ***\\
    \texttt{Qwen2.5-72B~\cite{qwen2.5}} & 86.67\% & 54.00\% & 20.00\% & 61.11\% & 86.67\% & 62.00\% & 40.00\% & 67.78\% & ***\\
    \texttt{LLaMA3.2-Vision-90B~\cite{llama3.2}}  & 63.33\% & 66.00\% & 10.00\% & 58.89\% & 66.67\% & 74.00\% & 30.00\% & 66.67\% & ***\\
    
    \hline
    \textbf{Proprietary tools and models} & & & & & & & & & \\
    \hline
    \texttt{Copilot Workspace~\cite{copilotworkspace}} & 80.00\% & 50.00\% & 30.00\% & 57.77\% & 86.67\% & 70.00\% & 50.00\% & 73.33\% & **\\
    \texttt{OpenAI-o1-2024-12-17~\cite{GPTo1}}  & 80.00\% & 62.00\% & 60.00\% & 67.78\% & 93.33\% & 74.00\% & 80.00\% & 81.11\%  & *\\
    \texttt{OpenAI-o3-mini-2025-01-31~\cite{GPTo3}}  & 83.33\% & 54.00\% & 50.00\% & 63.33\% & 90.00\% & 72.00\% & 60.00\% & 76.78\% & ***\\
    \texttt{Claude3.7-sonnet~\cite{claude3.7}} & 76.67\% & 62.00\% & 40.00\% & 64.44\% & 80.00\% & 74.00\% & 70.00\% & 75.56\% & ***\\
    \texttt{DeepSeek-R1~\cite{deepseekai2025deepseekr1incentivizingreasoningcapability}} & 83.33\% & 72.00\% & 50.00\% & 73.33\% & 93.33\% & 88.00\% & 80.00\% & 88.89\% & n.s.\\
    
    \hline
    \textbf{Multi-agent frameworks using OpenAI API} & & & & & & & & & \\
    \hline
    \texttt{ChatDev with GPT-3.5 (default)~\cite{qian2024chatdev}} & 53.33\% & 24.00\% & 0.00\% & 31.11\% & 70.00\% & 34.00\% & 20.00\% & 44.44\% & ***\\
    \texttt{Chatdev with GPT-4o~\cite{qian2024chatdev}} & 70.00\% & 54.00\% & 30.00\% & 56.67\% & 83.33\% & 74.00\% & 60.00\% & 75.56\% & ***\\
    \texttt{MetaGPT with GPT-4o~\cite{hong2023metagpt}} & 53.33\% & 44.00\% & 20.00\% & 44.44\% & 60.00\% & 52.00\% & 30.00\% & 52.22\% & ***\\
    \hline
    \textbf{Ours} & & & & & & & & & \\
    \hline
   
    \texttt{MaCTG without multi-scale evaluation} & 73.33\% & 56.00\% & 20.00\% & 57.77\% & 83.33\% & 68.00\% & 30.00\% & 68.89\% & ***\\
    \texttt{MaCTG without Think-twice extension} & 86.67\% & 52.00\% & 30.00\% & 61.11\% & 93.33\% & 68.00\% & 50.00\% & 74.44\% & ***\\
    \texttt{MaCTG without context-aware planning adjustment} & 76.67\% & 62.00\% & 30.00\% & 63.33\% & 83.33\% & 66.00\% & 40.00\% & 68.89\% & ***\\
    \texttt{MaCTG without any graph-based message passing} & 76.67\% & 44.00\% & 20.00\% & 52.22\% & 83.33\% & 60.00\% & 30.00\% & 64.44\% & ***\\
    \texttt{MaCTG with independent reasoning replacing CTG} & 70.00\% & 32.00\% & 30.00\% & 44.44\% & 76.67\% & 40.00\% & 30.00\% & 51.11\% & ***\\
    \texttt{MaCTG (with DeepSeek-V3 \& Qwen2.5-Coder-7B)} & \textbf{90.00\%} & \textbf{78.00\%} & \textbf{80.00\%} & \textbf{83.33\%} & \textbf{100.00\%} & \textbf{92.00\%} & \textbf{90.00\%} & \textbf{94.44\%} & --\\
    \hline
  \end{tabular}
  \footnotesize
  Statistical significance (Sig.) vs. MaCTG on Pass@5 using McNemar's test: ***: p $<$ 0.001, **: p $<$ 0.01, *: p $<$ 0.05, n.s.: not significant, --: reference method.
  \end{center}
\end{table*}

\subsubsection{Performance of MaCTG}
Table~\ref{tab:comparison} presents the performance of various automatic programming tools on the BCVPP dataset, with the last row reporting the results of MaCTG with all functionalities fully activated.  
For \emph{simple} tasks, which typically require only one module, the graph structure of MaCTG naturally converges into a sparse tree or even a linear chain structure.  
In these cases, the impact of MaCTG’s planning optimisation strategy is relatively limited; however, its multi-scale verification mechanism remains highly effective in reducing code errors.  
As a result, MaCTG achieves the highest accuracy (90.00\% Pass@1, 100.00\% Pass@5) in this category, demonstrating robust performance across multiple generation attempts.
For \emph{medium}-difficulty tasks, MaCTG’s planning and evaluation strategies are fully leveraged.  
Planner agents ensure a precise decomposition from modules to functions, allowing Executor agents to implement code with greater reliability.  
MaCTG attains strong performance (78.00\% Pass@1, 92.00\% Pass@5), demonstrating the effectiveness of its structured multi-agent workflow in handling moderate complexity.
For \emph{hard} tasks, where problem statements lack explicit guidance, MaCTG’s collaborative reasoning strategy proves advantageous. 
Its systematic planning capability, supported by inter-hierarchy and peer-agent collaboration, enables it to maintain robust task structuring even in complex scenarios. 
MaCTG achieves impressive results (80.00\% Pass@1, 90.00\% Pass@5), underscoring its adaptability in challenging programming tasks where multiple generation attempts reveal the framework's underlying capability.

\noindent\fbox{%
    \begin{minipage}{\dimexpr\linewidth-2\fboxsep-2\fboxrule}
        \textbf{Answer to RQ1}: MaCTG exhibits strong adaptability in auto-programming tasks.  
        Its collaborative reasoning strategies improve subtask decomposition, and the multi-scale validation mechanism ensures high coding accuracy, achieving an overall 94.44\% accuracy in handling auto-programming challenges.
    \end{minipage}%
}

In comparison with other auto-programming frameworks, large open-source models achieve moderate performance, with most reaching 50-60\% Pass@1 accuracy and 60-70\% Pass@5 accuracy.  
An exception is CodeLLaMA-Instruct~\cite{roziere2023code}, which, due to its older architecture and lower-quality training data, lags behind its counterparts. 
Notably, LLaMa3.3~\cite{llama3.1}, Qwen2.5~\cite{qwen2.5} and LLaMA-3.2-Vision~\cite{dubey2024llama} all reach around 60\% Pass@1 accuracy, with Pass@5 results showing marginal improvement (60-67\%), indicating performance saturation. 
These models perform well on simple and medium tasks but struggle with more complex problems, suggesting that even relatively large-scale models lack an effective task decomposition mechanism, limiting their ability to break down complex programming challenges.  
These observations further highlight the advantages of MaCTG’s Plan\&Execute workflow, which is specifically designed to address such difficulties through structured task decomposition and adaptive execution.
Proprietary tools and models clearly demonstrate the strongest overall performance among all baselines, with most achieving 60-75\% Pass@1 and 75-90\% Pass@5 accuracy.
The superior scale and quality of their training data contribute to their success across all difficulty levels.  
In contrast, multi-agent frameworks, ChatDev~\cite{qian2024chatdev} and MetaGPT~\cite{hong2023metagpt}, which rely on the OpenAI API, exhibit weaker results, primarily because they assign all tasks within a stage to a single agent, making complex contexts overly challenging and increasing the likelihood of hallucinations.
Statistical analysis confirms that MaCTG's improvements over these baselines are highly significant for most comparisons, with only DeepSeek-R1 showing non-significant differences (p = 0.344), indicating its genuine strength as a baseline. This validates that MaCTG's superior performance results from architectural advantages rather than random variation.

\begin{figure}[htbp]
    \centering
    \includegraphics[width=\linewidth]{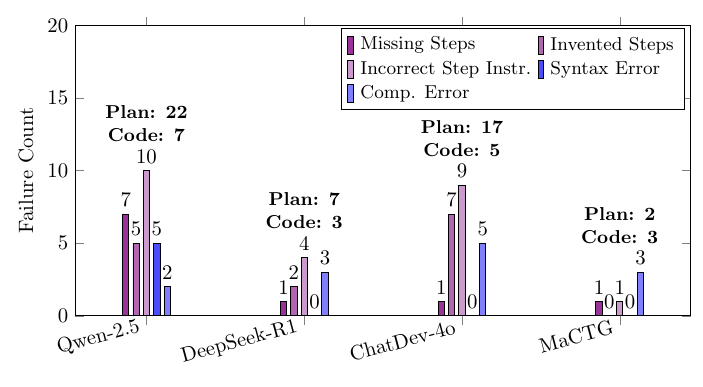}
    \caption{Failure statistics by error type. Missing Steps, Invented Steps, and Incorrect Step Instr represent planning failures (Plan), while Syntax and Comp Errors are coding failures (Code).}
    \label{fig:failure_bar}
    
\end{figure}

To quantify MaCTG's effectiveness in reducing hallucination, we conducted a failure case analysis comparing the best-performing method in each category to MaCTG.
Failure causes were grouped into five categories: 1) missing steps, 2) invented steps, and 3) incorrect step instructions represent hallucinations during the planning phase, where models fail to adhere to specified requirements, while 4) syntax and 5) computational errors represent hallucinations during the coding phase, where models generate plausible but incorrect code implementations. 
The failure distribution is summarised in Figure~\ref{fig:failure_bar}, where $Plan$ and $Code$ indicate the total number of planning and coding phase failures, respectively.
MaCTG demonstrates superior overall accuracy and remarkable reduction of hallucinations in both phases.
This demonstrates the effectiveness of the CTG framework in reducing planning-phase hallucinations and the multi-scale validation in preventing coding-phase hallucinations, providing comprehensive error mitigation throughout the automatic programming workflow.

\noindent\fbox{%
    \begin{minipage}{\dimexpr\linewidth-2\fboxsep-2\fboxrule}
        \textbf{Answer to RQ2}: MaCTG outperforms both open-source and proprietary methods across various difficulties. While open-source showed limitations on difficult problems, MaCTG’s planning strategy and evaluation process proved highly effective. Compared to other multi-agent frameworks, MaCTG achieved superior results in decomposing complex tasks. Additionally, failure analysis reveals that MaCTG significantly reduces hallucination errors, demonstrating more reliable code generation.
    \end{minipage}%
}

\subsubsection{Ablation study}
To assess the contributions of MaCTG's key architectural components, we conducted an ablation study by systematically disabling individual mechanisms.

\textbf{Multi-scale validation} showed significant impact across all difficulty levels, especially on hard tasks (90\% → 30\%), with overall Pass@5 accuracy dropping from 94.44\% to 68.89\%.
This mechanism plays a crucial role in ensuring execution reliability by systematically detecting and correcting errors before they propagate through the system.
Removing \textbf{Think-twice extension} caused moderate degradation (94.44\% → 74.44\% overall Pass@5). 
Without it, agents proceed with incomplete plans, leading to insufficient task decomposition and suboptimal code structure.
Disabling \textbf{Context-aware planning adjustment} reduced overall Pass@5 performance from 94.44\% to 68.89\%, with significant drops on medium (92\% → 66\%) and hard tasks (90\% → 40\%).
This mechanism prevents inconsistencies between functions and modules, ensuring agents operate with awareness of each other's outputs.
\textbf{Collaborative Thought Graph framework} evaluation involved removing all graph-based message passing while retaining the hierarchical structure. 
Planners operating in isolation resulted in 30\% performance degradation, demonstrating that graph-based collaboration is essential for preventing hallucinations and maintaining coordination.
\noindent\fbox{%
    \begin{minipage}{\dimexpr\linewidth-2\fboxsep-2\fboxrule}
        \textbf{Answer to RQ3}: The ablation study reveals that multi-scale validation systematically prevents error propagation, while CTG ensures inter-module consistency. Their combination demonstrates that structured collaboration with runtime validation is essential for reliable automatic programming.
    \end{minipage}%
}
    
We further evaluated collaborative versus independent reasoning by replacing Planner agents with individual reasoning models and removing graph-based message passing, creating a framework where each agent operates in isolation.
Results in the second-to-last row of Table~\ref{tab:comparison} show substantial performance degradation (94.44\% → 51.11\% overall Pass@5), demonstrating that independent reasoning leads to critical coordination failures.
Agents operating independently lack awareness of peer activities, resulting in incorrect cross-module references.

\vspace{3pt}
\noindent\fbox{%
    \begin{minipage}{\dimexpr\linewidth-2\fboxsep-2\fboxrule}
        \textbf{Answer to RQ4}: MaCTG’s collaborative reasoning offers advantages over independent reasoning by mitigating inconsistencies and enhancing global task coherence. Through collaboration, agents maintain interdependent awareness, ensuring logical consistency and seamless integration across modules.
    \end{minipage}%
}

\subsubsection{Cost-efficiency analysis}
\begin{figure}[htbp]
    \centering
     \includegraphics[width=\linewidth]{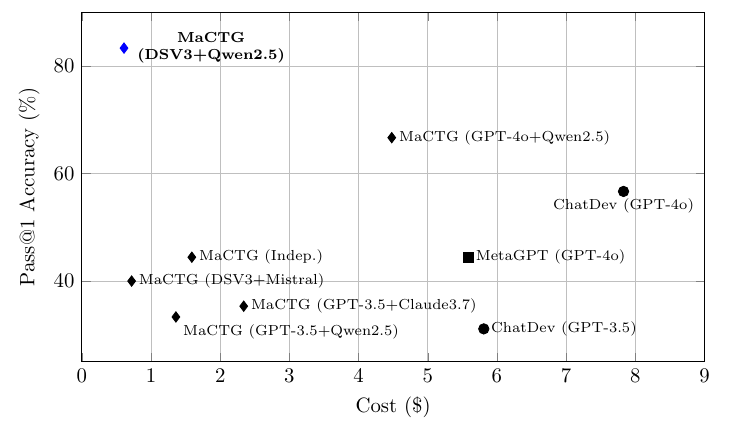}
    \caption{Cost-efficiency comparison of multi-agent systems.}
    \label{fig:cost-acc-scatter}
\end{figure}

A key limitation of multi-agent systems is their cost, particularly when relying on proprietary models.
To provide a more intuitive comparison of the trade-off between cost and accuracy, we visualise each method as a point in Figure~\ref{fig:cost-acc-scatter}.  
Among these frameworks, ChatDev incurred the highest cost due to redundant inter-agent communication, exceeding \$5 even with GPT-3.5, with costs increasing by over 34\% when using GPT-4o.  
MetaGPT reduced costs compared to ChatDev through refined agent roles and optimised prompts, yet its cost remained high.  

To comprehensively evaluate cost-efficiency, we tested MaCTG under various hybrid configurations, combining different proprietary and open-source LLMs for Planner and Executor agents, including GPT-4o+Qwen2.5-7B, DeepSeek-V3+Mistral-7B, GPT-3.5+Claude3.7 and GPT-3.5+Qwen2.5-7B. 
Using independent reasoning incurs more reasoning token consumption due to overthinking, even when there are fewer interactions among agents.
Leveraging collaborative reasoning and an optimal mix of models, MaCTG achieved an 89.09\% reduction in cost compared to standard multi-agent settings, while maintaining high accuracy.
\vspace{3pt}
\noindent\fbox{%
    \begin{minipage}{\dimexpr\linewidth-2\fboxsep-2\fboxrule}
        \textbf{Answer to RQ5}: MaCTG's collaboration strategy combined with hybrid configuration improves cost efficiency by assigning complex planning tasks to different proprietary models for better reasoning and basic coding tasks to open-source models to reduce expenses.
        This approach greatly reduces the costs compared with other multi-agent frameworks.
    \end{minipage}%
}

\subsection{Reasoning Trace Analysis}
To provide insights into how MaCTG's collaborative framework reduces hallucinations in practice, we present a detailed reasoning trace from a representative hard-level project in our BCVPP dataset. 

\textbf{Case Study - License Plate Recognition System.} 
Typically, we examine this project requiring three modules: image preprocessing, plate detection, and character recognition.
This case illustrates how MaCTG's collaborative reasoning prevents algorithmic hallucination that commonly occurs in multi-agent systems.

\textbf{Baseline Failure (ChatDev-4o):} ChatDev's Programmer agent, when tasked with implementing the detection module, stated: \textit{"For simplicity, let's assume the license plate is at a fixed position"} and implemented hard-coded cropping coordinates instead of actual detection algorithms.
Since ChatDev lacks mechanisms for other agents to review and challenge this oversimplified assumption, the Programmer agent directly began coding based on this incorrect plan without validation from peers or supervisors.

\textbf{MaCTG Correction Process:} \textit{Step 1 - Think-Twice Planning:} The Module Leader for plate detection initially considers various approaches, including Hough line detection and contour detection, before settling on a preliminary plan.
\textit{Step 2 - Context Sharing:} Peer Module Leaders exchange functional requirements: ML1 (Preprocessing): "Outputs Gaussian filtered greyscale image for robust contour detection", ML3 (Recognition): "Requires accurately cropped plate regions with clear texts". Upon receiving these specifications, the detection ML recognises that the preprocessing module is designed to support contour detection, and the recognition module needs an accurate plate region.
Therefore, the fixed-position approach would violate both the preprocessing design and recognition requirements. The Module Leader then commits to contour-based detection to properly utilise the inter-module coordination.
\textit{Step 3 - Supervised Adjustment:} Team Leader validates the overall workflow: "Detection module correctly utilises preprocessing output for dynamic plate localisation - approved."

\textbf{Result:} MaCTG generated a complete implementation using OpenCV's contour detection for robust plate localisation, while ChatDev produced a hallucinated non-functional system that only works for images with plates at predetermined positions.

\subsection{Threats to validity}
Despite its advantages, MaCTG has certain limitations:
1) Execution Speed: Reliance on locally deployed open-source models for Executors limits inference speed, and multi-scale validation involving iterative refinement increases runtime compared to single-model approaches. However, execution time remains comparable to ChatDev and faster than independent reasoning.
2) Stability: The dynamic agent structure introduces planning variability, as collaborative reasoning may lead to minor variations in generated plans. 
3) Generalisation: MaCTG currently focuses on image processing tasks. While the framework is generalisable, extending to other domains may require minor modifications to prompts, execution tools, and validation procedures.

\section{Conclusion}
In this work, we introduce MaCTG, a graph-based collaborative multi-agent framework for automated programming.
By incorporating context-aware planning adjustment and a multi-scale validation, MaCTG effectively reduces hallucinations and ensures structured project-level development.
Its hybrid LLM configuration optimises computational cost by strategically allocating tasks between proprietary and open-source models.
Experimental evaluations demonstrate that MaCTG outperforms other methods, achieving higher accuracy across varying task complexities.

Future research will focus on optimising execution efficiency and improving stability to further advance MaCTG’s applicability in real-world auto-programming.
With continued refinement, the collaborative thought graph framework has the potential to tackle even more complex programming challenges, supporting developers and researchers in building reliable automated solutions.

\bibliographystyle{ACM-Reference-Format}
\bibliography{sample-base}

\end{document}